%% file: Paper.tex
\documentclass[12pt]{iopart}

\def\bfu{\mbox{\bf u}}
\def\bfB{\mbox{\bf B}}

\def\T{T}
\def\bfe{\mbox{\bf e}}
\def\bfr{\mbox{\bf r}}

\def\bfnabla{\mbox{\boldmath $\nabla$}}
\def\bfOmega{\mbox{\boldmath $\Omega$}}
\newcommand*{\Ray}{{\rm Ra}}

\newcommand*{\Pm}{{\rm Pm}}

\newcommand*{\Rm}{{\rm Rm}}
\newcommand*{\q}{{\rm q}}
\newcommand*{\Ek}{{\rm E}}
\newcommand*{\Nu}{{\rm Nu}}

\newcommand{\Ro}{{\rm Ro}}

\newcommand{\B}{{\rm B}}

\usepackage{harvard} 
\usepackage{color}
\usepackage{graphicx}

\begin{document}

\title[]{Three Branches of Dynamo Action}

\author{Emmanuel Dormy$^1$\footnote{Corresponding 
author: dormy@dma.ens.fr}, Ludivine Oruba$^{2,3}$ and Ludovic Petitdemange$^{2,3}$}

\address{$^1$Department of Mathematics \& Applications, CNRS UMR 8553, 
Ecole Normale Sup\'{e}rieure, Paris, France}
\address{$^2$Physics Department, Ecole Normale Sup\'{e}rieure, Paris, France}
\address{$^3$LERMA, CNRS UMR 8112, Observatoire de Paris, Paris, France}

\ead{dormy@dma.ens.fr}

\begin{abstract}

In addition to the weak-dipolar state and to the fluctuating-multipolar state, widely
discussed in the literature, a third regime has been identified in 
\cite{Dormy16}.  
It corresponds to a strong-dipolar branch which appears to approach, in a
numerically affordable regime, the magnetostrophic limit relevant to
the dynamics of the Earth's core.
We discuss {the transitions between these states and point to}
 the relevance to this strong-dipolar state to Geodynamo modelling.
\end{abstract}

\vspace{2pc}
\noindent{\it Keywords}: Dynamo action, Magnetostrophic balance, 
Dynamo bifurcation.

\maketitle

\section{Introduction}

The Earth's magnetic field is sustained by self-exciting dynamo action in the
liquid core of our planet. Part of the kinetic energy of the flow is
transferred to magnetic energy. In fact, in the Earth's core, most of the energy
is anticipated to be dissipated by electrical currents. We describe here
the existing numerical models and show how they appear to fall into three
distinct branches characteristic of different forces balances.

Since the first full numerical models of dynamo action
\cite{ZhangBussea,ZhangBusseb,Glatzmaier}, many 
parameter space surveys have been performed. This has allowed
to produce phase-diagrams which, depending on the controlling parameters,
describe which dynamo states can be achieved. Wide parameters surveys
\citeaffixed{Chr06,SPD12}{e.g.}  clearly identified two branches of dynamo
action. The first one is characterised by a dominant axial dipole, while
the second one, at larger forcing, is largely multipolar, with a fluctuating
dipolar component.
It was shown \cite{SimitevBusse} that the transition between these two dynamo
states can be hysteretic if stress-free boundary conditions were
considered.

Sadly, upon closer investigation none of these two
states turned out to be relevant for the Geodynamo. 
The first state, characterised by a dominant axial dipole, was shown to be
largely controlled by viscous effects \cite{King13,Oruba14GJI}, and 
the second one to involve significant inertial effects 
\cite{Chr06,SPD12,Oruba14GRL}.

The Earth rotates with one revolution per day, given the viscosity of
liquid iron at these temperature and pressure \citeaffixed{deWijs}{see}, we can
easily conclude that viscous effects will only become relevant at very small
scales. A naive dimensional analysis approach yields $\ell_\nu ^2 \sim \nu
/ \Omega$. Asymptotic developments \citeaffixed{DormySoward}{e.g.} 
reveal more elaborate scalings of the
form $\ell_\nu ^\alpha \sim \nu
/ \Omega \, L^{\alpha -2}$ (where $L$ denotes the typical size of the Earth's core). 
The first relevant length scale for a vertical shear, for example,
corresponds to $\alpha = 3$~.
For geophysically relevant estimates, these length scales would be of a few
meters, less than a kilometer. Length scales which lie below the resolution of current
numerical models. 

Typical velocities in the Earth's core can be inferred from the secular
variation of the magnetic field \cite{Holme07}, this yields $U \simeq 10^{-4}{\rm
  m} \,{\rm s}^{-1} \, .$ There again, dimensional analysis reveals the
length scale at which inertial effects will be comparable to the effects of
global rotation (the so-called Rossby radius), $\ell_{I} \sim U/\Omega$,
again of the order of a few meters. So that inertial effects are not
expected to play a significant role, on the time scale of secular
variation, at the large spatial scale.

The relevant balance for the Earth's core is therefore one in which both
viscous effects and inertial effects are negligible on the large
scales. This is known as the magnetostrophic balance \citeaffixed{Moffatt}{see}.
The issue of whether the limit system of equations (i.e. omitting both the viscous term
and the inertial term in the governing equations) is well posed is a challenging one. 
First put forward by Taylor in 1963 \cite{Taylor63}, this system proved
extremely difficult to solve numerically. The well-posedness of this limit
raises complicated mathematical issues \cite{Gallag16}. A first set of
solutions in the case of an axi-symmetric configuration has however
recently been achieved \cite{RobertsWu,WuRoberts}. 

We here take the simpler point of view of retaining all the terms in the
equations, but vary the parameters, so as to try and approach a
magnetostrophic equilibrium.  The crucial issue in doing so, is to assess
that inertial and viscous effects are indeed small in the realised
solution. To that end, a useful tool is to study the bifurcation diagram
for various choices of the parameters, the different branches on this
diagram naturally corresponding to different forces balances.  The
bifurcation diagram for dynamo action in the limit relevant to the Earth's
core has been the focus of many analytical or mixed analytical-numerical
studies \cite{EltayebRoberts,ChildressSoward,Soward74,Fautrelle,Roberts88}.
The main result is that the bifurcation diagram should consist of two
branches. The first branch necessarily involves significant viscous effects
and is referred to as the ``weak-field'' branch.  On this branch, viscous
effects are necessary to allow deviation from the Proudman-Taylor
constraint (i.e. the tendency for the flow in a rapidly rotating reference
frame to be independent on the coordinate in the direction of the axis of
rotation). The flow will thus develop short length scales in the directions
orthogonal to the axis of rotation. The most obvious of these length scales
involving an $\Ek^{1/3}$ dependence \citeaffixed[for
  example]{DormySoward}{see}. As the strength of convection increases, both
the flow and the field gain in amplitude. A transition, characterised by a
turning point, is anticipated when the Lorentz force becomes large enough.
When this turning point is reached, the weak-field solution becomes
unstable, and the magnetic field experiences a runaway
amplification. Saturation will be achieved when the field reaches a
strength sufficient for the Lorentz force to be comparable with the
Coriolis force. This second branch is referred to as the ``strong-field''
branch. On this branch the amplitude of the Lorentz force is comparable to
that of the Coriolis force.

The above description was so far disconnected from direct numerical
simulations of spherical dynamos.  Recently, however, \citeasnoun{Dormy16}
pointed out the existence of a third dynamo state, numerically achievable
at the cost of an under-estimated magnetic diffusivity. This regime appears
to approach the relevant magnetostrophic force balance.  This
strong-dipolar dynamo state, described numerically, is characterised by an
hysteresis with respect to the viscous-state. The transition occurs at a
turning point, which is characterised by a runaway field growth. This
bifurcation sequence establishes a first connection between direct
numerical models and earlier asymptotic developments.

In this article, we will rapidly review the available results on
dynamo states available from numerical simulations. We further investigate
the numerical strong-dipolar (SD) branch, describe its relation with the
earlier theoretical bifurcation sequence, and ponder on the relevance of
the strong-dipolar state to the Geodynamo.

\section{Governing equations}
Let us start by introducing the standard mathematical model for the Geodynamo.
The numerical simulations discussed in this paper are restricted Boussinesq models.
The computational domain consists of a spherical shell with
aspect ratio $ r_i/r_o=0.35 \, . $
The flow is thermally driven, and a fixed difference of temperature is imposed between the inner 
and outer boundaries. It should be noted, that in the Earth's core, buoyancy effects are associated 
with both  thermal and compositional effects. The simplest form of governing equations is however similar 
in both cases 
\citeaffixed[for a detailed discussion]{Braginsky}{see}.
All the simulations used in this work rely on no-slip mechanical boundary
conditions as well as an insulating outer domain. The inner core is
insulating in most simulations, and a few simulations involve a conducting inner core 
with the same conductivity as the fluid.

The governing equations in the rotating frame of reference can then be
written in their non-dimensional form -- using $L=r_o-r_i$ as unit of length, $L^2/\eta$ as unit of
time, $\Delta T$ as unit of temperature, and 
$\left({\rho \mu \eta \Omega}\right)^{1/2}$ as
unit for the magnetic field -- as
\begin{equation}
{\Ek}_\eta
\left[\partial _t \bfu +  (\bfu \cdot \bfnabla) \bfu 
\right]=
- \bfnabla \pi 
\,+\, {\rm E} \, \Delta \bfu
\,-\, 2 \bfe_z \times \bfu 
\,+\, {\Ray}\,\q\, \T \, \bfr
\,+\, \left(\bfnabla \times \bfB \right) \times \bfB\, , \,\,\,\,\,\,
\label{eq_NS}
\end{equation} 
\begin{equation}
\partial _t \bfB = \bfnabla \times (\bfu \times \bfB) 
+ \Delta \bfB \, ,
\qquad
\partial_t \T + (\bfu \cdot \bfnabla) \T
= \q \,\Delta \T\, ,
\label{eq_ind}
\end{equation}
\begin{equation}
{\rm with} \qquad \bfnabla \cdot \bfu = 
\bfnabla \cdot \bfB = 0\, .
\label{eq_div}
\end{equation} 
System~(\ref{eq_NS}--\ref{eq_div})
involves four independent non-dimensional parameters, which are 
the Ekman number $\Ek = \nu / (\Omega L ^2) \, ,$
the magnetic Ekman number $\Ek_\eta = \eta / (\Omega L ^2) \, ,$
the Roberts number ${\rm q}={\kappa}/{\eta}\, , $ 
and the modified Rayleigh number ${\Ray} = \alpha g \Delta T L / (\kappa \Omega)\, ,$
in which $\nu$ is the kinematic viscosity of the fluid, 
$\alpha$ the coefficient of thermal expansion, $g$ is  
the gravity at the outer bounding sphere (the gravity profile is linear in radius), 
$\kappa$ its thermal diffusivity, and $\eta$ its magnetic diffusivity.
The modified Rayleigh number ${\Ray}\, ,$ as defined above, differs from its most 
classical definition $\alpha g \Delta T L^3 / (\nu \kappa) \, ,$ to which it is
related via an Ekman factor.
Whereas the later is the relevant parameter to measure energy input in the
standard 
Rayleigh-B\'enard setup, 
it is not any longer relevant in the magnetostrophic limit.

The above four independent parameters are enough to fully define the system. 
{We shall now consider}
 the relevant values of these parameters for the Earth's core.
The orders of magnitude of the dimensional coefficients outlined in the introduction reveal
\begin{equation}
\Ek \simeq 10^{-15} \, , \qquad
\Ek_\eta \simeq 10^{-9} \, , \qquad
{\rm q} \simeq 10^{-5} \, .
\label{small_param}
\end{equation}
The last non-dimensional parameter, ${\Ray}$, controlling
the strength of thermal convection is difficult to quantify in a
Boussinesq formalism. While the heat gradient across the core is of 
the order of $10^3$~K, most of the heat in the actual core
is carried along the adiabat. 
Only the super-adiabatic gradient is
relevant in the Boussinesq framework. This deviation
is only of the order of $10^{-3}$K \citeaffixed{Gubbins01,JonesRev}{see},
so difficult to estimate with great precision.
This results in the following estimate for our definition of the Rayleigh number
$\Ray \simeq 10^{13} $ \cite{Gubbins01}.
Obviously this value should be large enough, so that, even though $\q$ is a
vanishing number, the product $\widetilde{\Ray} = \Ray \,\q \, $ remains of order unity.
The above estimate yields
$\widetilde{\Ray} \simeq 10^8 \, .$

The non-dimensional form chosen in~(\ref{eq_NS}--\ref{eq_div}), often referred to as the 
``strong-field scaling'', 
highlights the primary magnetostrophic balance (order one terms) and the three vanishing parameters
(\ref{small_param}). The Taylor state introduced above \cite{Taylor63} amounts to dropping all small terms in 
~(\ref{eq_NS}--\ref{eq_div}). The only parameter left to control this limit system is then the modified 
Rayleigh number $\widetilde{\Ray}\, .$ So that the strength of the magnetic field in this limit should 
depend on this sole parameter.

It is worth pondering on the ratios of the small terms in~(\ref{eq_NS}--\ref{eq_div}). 
The ratio of the Ekman number to the magnetic Ekman number defines the
magnetic Prandtl number ${\rm Pm}=\Ek/\Ek_\eta\, ,$ this defines the
ratio of two small parameters (vanishing in the geophysically relevant
magnetostrophic limit), but as we shall see controlling this ratio in
the limiting process is essential.  The ratio of the magnetic Prandtl
number to the Roberts number defines the classical hydrodynamic
Prandtl number, $\Pr = \Pm / {\rm q} $. There again, both $\Pm$ and ${\rm q}$
are small numbers, but their ratio remains an important quantity.

The importance of the ratio of small parameters in a double (or even tripple in our case) limit, 
will not come as a surprise to the mathematical community. 
Besides, its importance has recently been stressed in physical applications when 
considering the 
saturation properties of MHD turbulence \cite{FromangPapa,Fromang}.

Our analysis is tested against a wide numerical database corresponding 
to some $300$ direct numerical simulations. The data sample is composed of $180$
runs kindly provided by U.~Christensen, and of additional runs, 
{either previously reported in \citeasnoun{Morin09}, 
\citeasnoun{SPD12} and \citeasnoun{Dormy16}, or presented in Table~1.}

\section{The Weak-Dipolar dynamo state}

The first and most documented dynamo state is for obvious reason the
dipolar state. Dynamos in this state have been reported since the very
early days of dynamo modelling. Owing to their dipolar nature, these
models have even often been argued to be relevant to the Geodynamo. It is
now evident that viscous effects are present at leading order in the forces
balances. 

\begin{figure}
\centerline{\includegraphics[width=0.45 \textwidth]{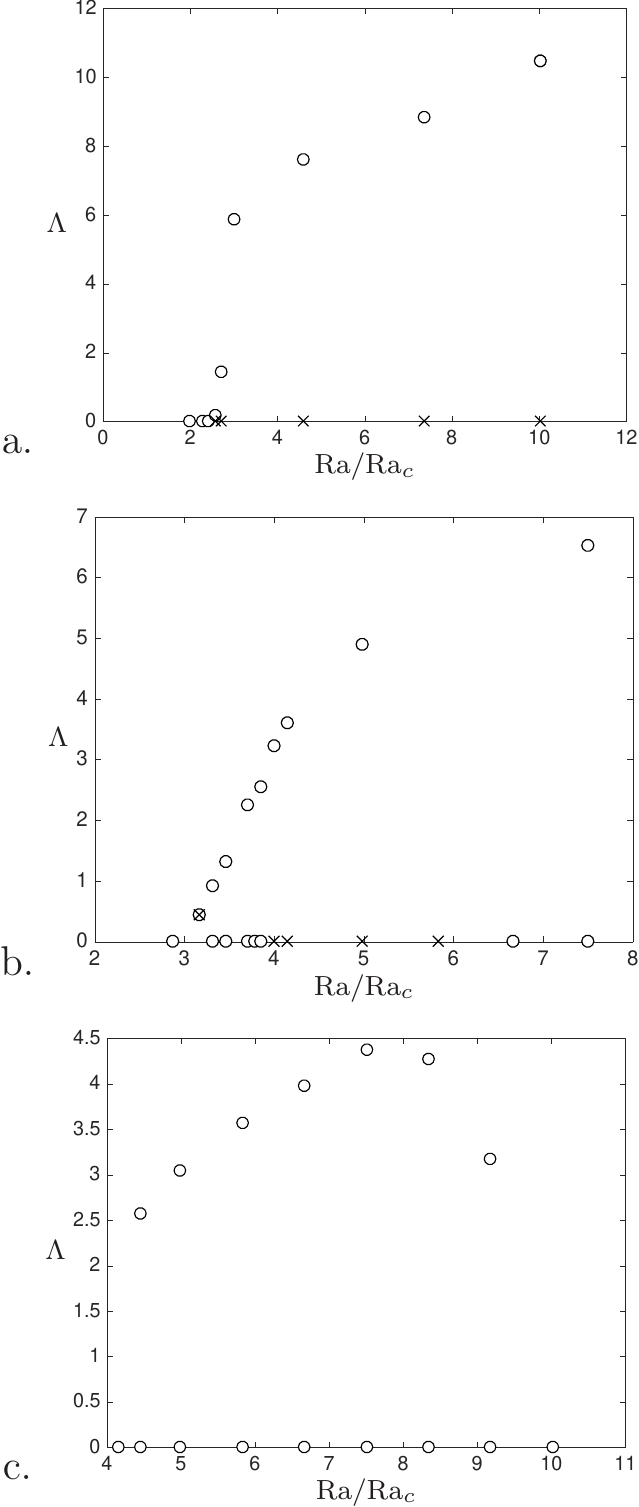}}
\caption{Bifurcation diagrams obtained for $\Ek=3\times 10^{-4}$ and 
(a)  $\Ek_\eta=5 \times 10^{-5} \, ,$ 
(b)  $\Ek_\eta=10^{-4} \, ,$ 
(c)  $\Ek_\eta=2 \times 10^{-4} \, $ ($\circ$ stable, 
    $\times$ unstable, ${\scriptstyle\otimes}$ meta-stable). 
  These respectively correspond to a
  super-critical, a sub-critical and an isola bifurcation diagram. The same
  sequence of transitions between the various bifurcation diagrams 
  were obtained at different values of $\Ek$, for similar values of $\Ek_\eta$
\citeaffixed{Morin09}{see}.}
\label{fig_Morin}
\end{figure}

This branch is the first dynamo mode produced as the modified Rayleigh
$\widetilde{\Ray}$ is increased away from its critical value for the onset
of convection $\widetilde{\Ray}_c$ (which can itself be subcritical 
\citeaffixed{Guervilly15}{see}).
The onset of dynamo action, or dynamo bifurcation, has been carefully
investigated in \citeasnoun{Morin09}. Depending on the parameters being
considered, they reported either super-critical, sub-critical or isola branches
for the onset of dynamo action (see figure~\ref{fig_Morin}). It is
important to stress that each point on these numerical bifurcation diagrams
corresponds to a time averaged fully three-dimensional simulation.

The Ekman number was varied between $10^{-3}$ and $10^{-4}$, and the
bifurcation type over this range appears to depend only on the parameter
$\Ek_\eta$.  Super-critical bifurcations were obtained for both 
$\Ek=3 \times 10^{-4}$ and $\Ek=10^{-4}$, with $E_\eta < 5 \times 10^{-5}$;
Sub-critical bifurcations for both $\Ek=3 \times 10^{-4}$ and $\Ek=10^{-4}$,
with $5 \times 10^{-5} < E_\eta < 2 \times10^{-4}$; and Isola were obtained for both
$\Ek=10^{-3}$ and $\Ek=3 \times 10^{-4}$ with $E_\eta > 2 \times 10^{-4}$. 
These simulations were all performed at fixed $\Pr = \Ek \, \Ek_\eta^{-1} \,
\q^{-1} = 1 \, .$

The physical explanation for this change of behaviour has not been
achieved so far, but the ordering highlighted above points to the
importance of inertial effects in controlling the nature of the 
transition.\footnote{Note that the original
paper \cite{Morin09} uses a different ordering based on the magnetic Prandtl number
at fixed Ekman number.}

The importance of the flow helicity on the dynamo generation mechanism for
this dynamo state has been highlighted by \citeasnoun{OlsonCG99}.
Besides, \citeasnoun{Sreeni11} 
argue that kinematic helicity enhancement by the
magnetic field could provide a mechanism for the occurrence of sub-critical 
dynamo branches.

The importance of viscous forces in this dynamo branch has long been
overlooked. It was however recently pointed out \cite{King13} that the
typical length scale of the flow exhibits a clear $\Ek^{1/3}$
dependence, characteristic as explained above of the viscous-Coriolis
dominant force balance.

\begin{figure}
\centerline{\includegraphics[width=1 \textwidth]{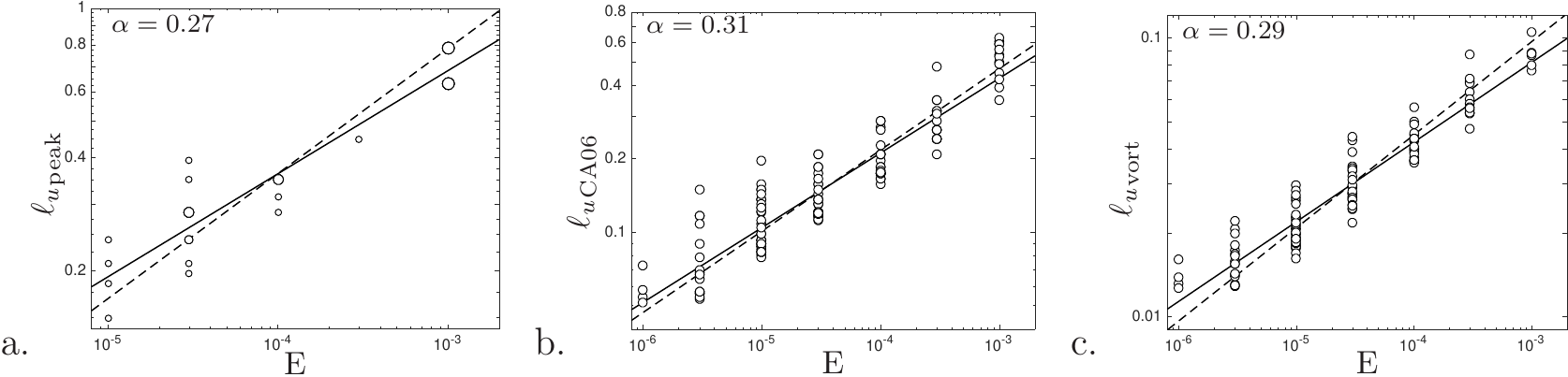}}
\caption{Evolution of the typical length scales of the flow as a function 
of the Ekman number for dynamos in the weak-dipolar state. 
(a) The length scale at which the energy spectrum peaks ${{\ell}_u}_{\rm peak}$ ;
(b) the averaged length scale ${{\ell}_u}_{\rm CA06}$; and (c) the vorticity length scale
${\ell_u}_{\rm vort}$.
The dashed line indicates the $\Ek^{1/3}$ scaling.
The best fit, derived from a least squares method, in $\Ek^{\alpha}$, 
is indicated by a solid line, and the corresponding value of $\alpha$ is 
reported on each panel. On the first graph, larger symbols are used when several dynamos produced the same ${{\ell}_u}_{\rm peak}\, .$}
\label{lu_weak}
\end{figure}

We present in figure~\ref{lu_weak} three different length scales.
The first one, ${{\ell}_u}_{\rm peak}$, is defined by considering the time averaged kinetic energy spectrum. 
It is defined as ${{\ell}_u}_{\rm peak}=\pi/ l_{\rm peak}$ where $l_{\rm peak}$ corresponds to the spherical 
harmonic degree for the peak of the energy spectrum. 
The second length scale ${{\ell}_u}_{\rm CA06}$ corresponds to the length scale defined in \citeasnoun{Chr06}
and used in \citeasnoun{King13}, it is defined as ${{\ell}_u}_{\rm CA06}=\pi/ l_{\rm CA06}$ where
$l_{\rm CA06}$  corresponds to the mean value
of the spherical harmonics degree in the time-averaged kinetic energy
spectrum \citeaffixed[equation (27)]{Chr06}{see}.
The third length scale ${\ell_u}_{\rm vort} \, ,$ introduced by \citeasnoun{Oruba14GJI}, is defined as
\begin{equation} 
{\ell_u}_{\rm vort} ^2 = \frac{\langle \mathbf{u}^2 \rangle}{\langle (\mathbf{\nabla} \times \mathbf{u})^2 \rangle} \, ,
\label{ldissu}
\end{equation} 
where $\langle \cdot \rangle$ denotes time and volume averaged quantities.

Figure~\ref{lu_weak} highlights that the three typical length scales defined above 
follow an Ekman dependence characteristic of the viscous, $\Ek ^{1/3}$, scaling.

Because viscous forces are important in this branch, and to attempt a link with the
earlier asymptotic studies listed above, we will in the sequel refer to this branch as the
weak-dipolar (WD) branch.
Of course this branch is saturated, and non-linear effects are affecting
the flow, both via the non-linear inertial term and via the Lorentz force.

\section{From Weak-Dipolar dynamos to Fluctuating-Multipolar dynamos}

As the forcing is increased, i.e. as the modified Rayleigh
${\Ray}$ is further increased away from ${\Ray}_c \, ,$ a transition to
a fluctuating-multipolar (FM) dynamo state has been initially reported by
\citeasnoun{Ku02} and described in further details in \citeasnoun{Chr06}.

This transition was very early associated with the strength of inertial
effects. Indeed \citeasnoun{Chr06} pointed out that the transition was 
controlled by 
the ``local'' Rossby number $\Ro_\ell = U (\Omega \, {{\ell}_u}^\star_{\rm CA06})^{-1}$
based on the mean velocity length scale ${{\ell}_u}_{\rm CA06} \, ,$ defined above
(the $\,^\star$ denotes dimensional quantities).
This clearly indicates that inertial effects become significant at the flow
length scale when the weak-dipolar mode is lost.
More recently, \citeasnoun{Oruba14GRL} showed that rather
than measuring the typical length scale of the realised flow, one could
account for the transition with the parameter $\Ro \, \Ek ^{-1/3}$, where 
$\Ro = U \, (\Omega L)^{-1}$, because of the above mentioned dependence of the flow
length scale as $\Ek^{1/3} \, L$ in the weak-dipolar branch.

\begin{figure}
\centerline{\includegraphics[width=1 \textwidth]{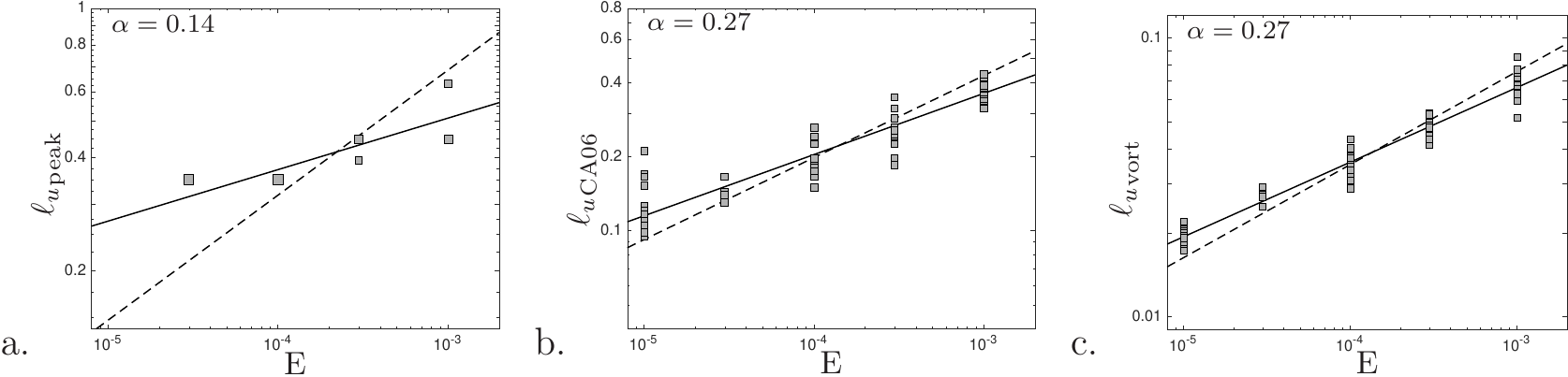}}
\caption{Evolution of the typical length scales of the flow as a function 
of the Ekman number, same representation as in figure~\ref{lu_weak}, but for numerical dynamos 
in the multipolar-fluctuating state.}
\label{lu_multi}
\end{figure}

The typical length scales are presented as a function of the Ekman number
on figure~\ref{lu_multi}.
 While the length scales ${\ell_u}_{\rm
  vort}$ and ${{\ell}_u}_{\rm CA06}$ are by construction affected by
the viscous dissipation length scale, the length scale
${{\ell}_u}_{\rm peak}$ (corresponding to the energy spectrum peak)
exhibits a much weaker dependence on the Ekman number (we stress however
the large dispersion due to the small data sample and the difficulty
to precisely estimate $l_{\rm peak}$, which is a small integer).
This appears as a signature of inertial effects at the dominant scale of the flow.

\begin{figure}
\centerline{\includegraphics[width=0.45 \textwidth]{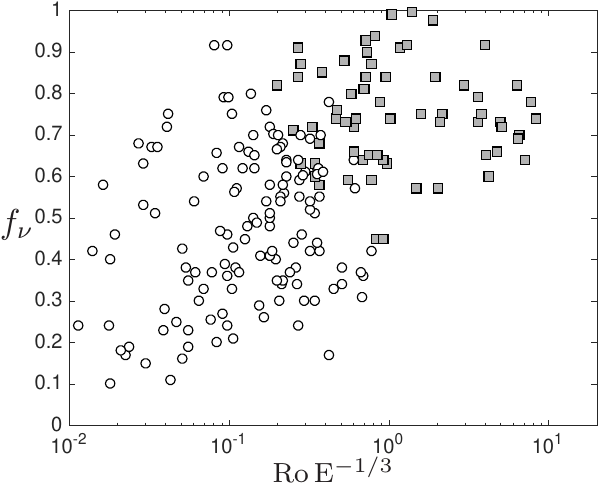}}
\caption{Fraction of viscous dissipation as a function of $\Ro \, \Ek ^{-1/3} \, .$
The transition from the weak-dipolar (circles) to the
fluctuating-multipolar state (grey squares) is clearly controlled by 
$\Ro \, \Ek ^{-1/3}$, characterising the importance of inertial effects at
the flow length scale. 
The fraction of viscous dissipation is significant in both regimes and
appears enhanced by inertial effects.} 
\label{transWM}
\end{figure}

A way to further assess the importance of viscosity is to 
{consider} the 
fraction of 
energy being dissipated by viscous forces $f_\nu$. The ratio of the energy
being dissipated by viscous forces to the total energy dissipation (viscous
and ohmic)
\begin{equation}
f_\nu=
 \frac{\Ek \langle (\mathbf{\nabla}\times \mathbf{u})^2 \rangle}
{\Ek \langle (\mathbf{\nabla}\times \mathbf{u})^2 \rangle +
\langle (\mathbf{\nabla}\times \mathbf{B})^2 \rangle
}
=
 \frac{\rho \nu \, \langle (\mathbf{\nabla}\times \mathbf{u}^{\star})^2 \rangle}
{\rho \nu \, \langle (\mathbf{\nabla}\times \mathbf{u}^{\star})^2 \rangle +
{\eta}{\mu}^{-1}\,
\langle (\mathbf{\nabla}\times \mathbf{B}^{\star})^2 \rangle
} 
\end{equation} 
is reported on figure~\ref{transWM} as a function of 
the Rossby number
based on the viscous scale \cite{Oruba14GRL}. The relevance of this
parameter $\Ro \, \Ek^{-1/3}$ to distinguish weak-dipolar and
fluctuating-multipolar dynamos is evident. 
We should stress that \citeasnoun{Schrinner13} investigated the behaviour
of $f_\nu$ (actually $f_{\rm ohm} = 1 - f_\nu $) and concluded that $f_\nu$ was increasing
with the local Rossby number,
for which $\Ro \, \Ek^{-1/3}$ offers a very good proxy in the weak-dipolar state. 

The fraction of viscously dissipated energy is in general significant
in both dynamo states (though some end members models of the
weak-dipolar state, reach 10\% of viscous dissipation). It is worth
noting that this fraction is on average larger in the multipolar
regime, for which the field is weaker and the ohmic dissipation is
thus lower.

\begin{figure}
\centerline{\includegraphics[width=0.45 \textwidth]{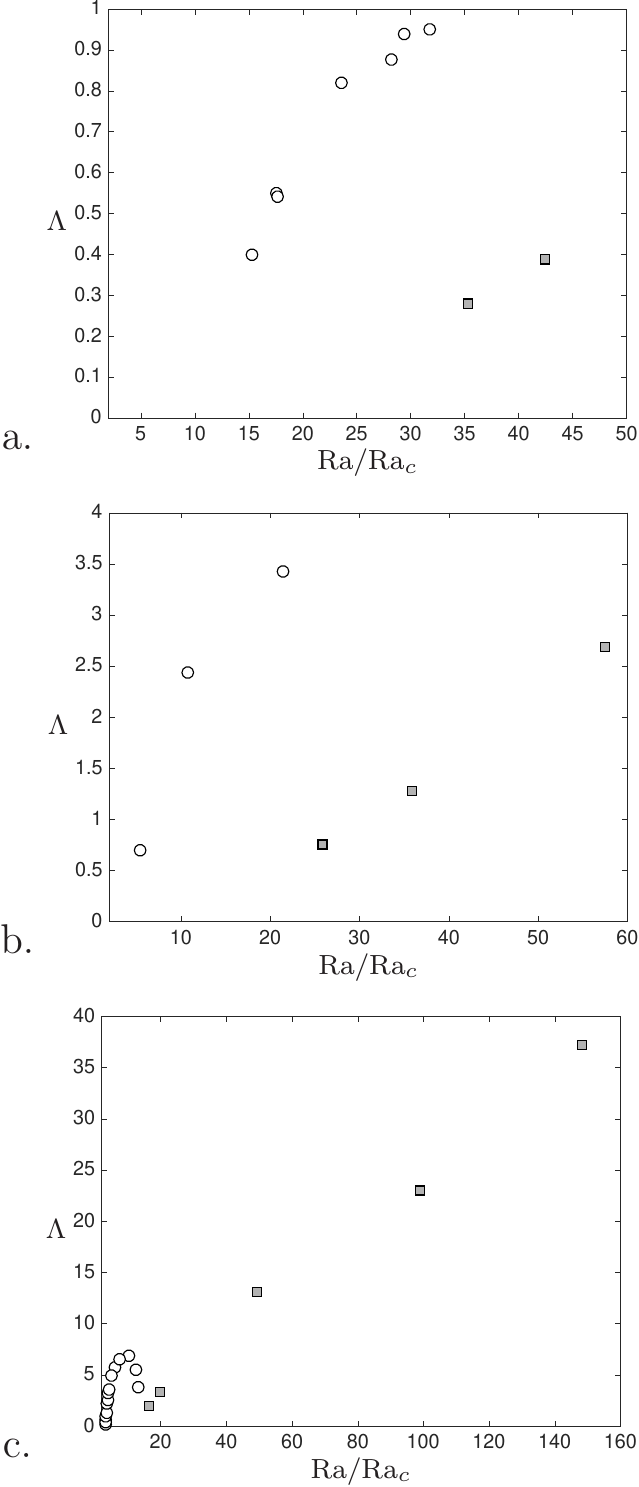}}
\caption{Weak-dipolar (circles) and fluctuating-multipolar (grey squares)
  branches 
with no-slip boundary conditions for 
(a) $\Ek=3 \times 10^{-5} \, ,$ $\Ek_\eta=1.2  \times 10^{-4} \, ,$
  $\q=0.25 \, ,$
(b) $\Ek=10^{-4}\, ,$ $\Ek_\eta=10^{-4}\, ,$ $\q=1 \, ,$
and (c) 
$\Ek=3 \times 10^{-4}\, ,$ $\Ek_\eta= 10^{-4}\, ,$ $\q=3 \, .$
}
\label{bif_multi1}
\end{figure}

The transition from the weak-dipolar to the fluctuating-multipolar state 
is illustrated in figure~\ref{bif_multi1}.
Here again the simulations were performed at fixed $\Pr = \Ek \,
\Ek_\eta^{-1} \, \q^{-1} = 1 \, .$ 
The two branches are identified with different symbols. These are easily
identified by measuring the strength of the dipolar component
relative to the total field intensity. The dipolar state is indicated
with circles, whereas the fluctuating-multipolar state is indicated with grey 
squares.
In these graphs the magnetic energy density of the non-dimensional 
magnetic field is reported: $\Lambda = \langle B^2 \rangle /2\,,$ 
this corresponds to 
the Elsasser number. In terms of dimensional variables, this
amounts to
\begin{equation}
\Lambda= \langle{\B^{\star}}^2\rangle/\left(2 \Omega \rho \mu \eta \right) \, .
\label{Elsasser}
\end{equation} 
Figures~\ref{bif_multi1}a and \ref{bif_multi1}b highlight the
discontinuity in the dynamo branches. Figure~\ref{bif_multi1}c shows that
for some parameters, the efficiency of the viscous dynamo state starts to
decrease before the transition occurs.
This highlights that although the primary force balance on this branch
involves the viscous term, non-linear effects are important, in saturating
the field growth, but also in modifying the flow (which, in this case, provides a lower
saturation level at larger forcing).

\begin{figure}
\centerline{\includegraphics[width=0.45 \textwidth]{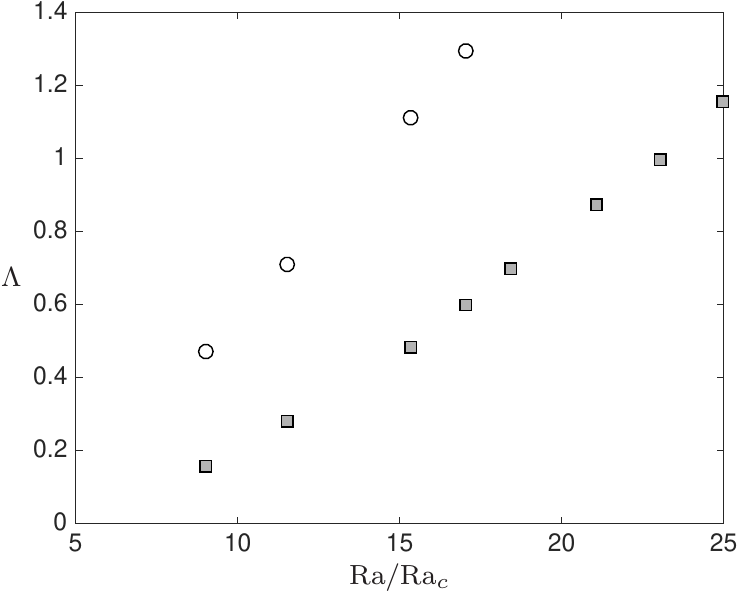}}
\caption{Weak-dipolar (circles) and fluctuating-multipolar (grey squares)
  branches for 
$\Ek=10^{-4} \, ,$ $\Ek_\eta=10^{-4} \, ,$ $\q=1 \, ,$ 
in the case of stress-free boundary conditions.}
\label{bif_multi2}
\end{figure}

Non-linear effects in a rotating flow are known to drive zonal flows
through the Reynolds stress. Such zonal flows are very weakly damped. If
their radial structure is larger than $\Ek^{1/4} L$ they are dominated by
boundary layers dissipation \cite{Morin06}.
It results that the large-scale zonal flows will behave in a different
manner in the case of stress-free boundary conditions (for which only the
bulk viscous effects will be relevant). 
This will, of course be particularly true when the Ekman number is
moderately small (as is the case in most numerical models). As the Ekman
number decreases, the correction due to the boundary layer dissipation (via
Ekman pumping) will become less and less important.
For stress-free boundary conditions the possible bistability between the
weak-dipolar state and the fluctuating-multipolar state was first highlighted by
\citeasnoun{SimitevBusse}.
This is directly associated with the zonal flow, which once present
prevents the formation of an organized large-scale field. So that the
transition to the fluctuating-multipolar branch is hysteretic and once on
this branch, the controlling parameter (the Rayleigh number) can be
decreased below the transitional value without recovering the weak-dipolar
state.
\citeasnoun{SPD12} further 
demonstrated that in this case the controlling parameter was still a local
Rossby number, but which needed to be based on the amplitude of the
convective flow, and not of the zonal flow itself.
This bistability is illustrated by figure~\ref{bif_multi2}.

This second dynamo state is characterised by fluctuations of the dipolar
component. For this reason, it has sometimes been argued that geomagnetic
polarity reversals may be due to the fact the Geodynamo operates near this
transition \cite{Olson06,Chr10}.
Indeed, an estimation of the parameter $\Ro \, \Ek ^{-1/3}$ in the Earth's
core would be very close to the critical value observed in numerical
simulations (below, but close to $10^{-1}$). 
It is however most unlikely that the Geodynamo operates on the viscous 
dipolar branch. The resulting viscous length scale would be extremely
small, less than $100$~m. The strong-dipolar state described previously is
most likely the relevant one, and this transition is probably not relevant
to the actual Geodynamo reversals
\citeaffixed{Oruba14GRL}{see discussions in}.

\section{The Strong-Dipolar dynamo state}

\begin{figure}
\centerline{\includegraphics[width=0.9 \textwidth]{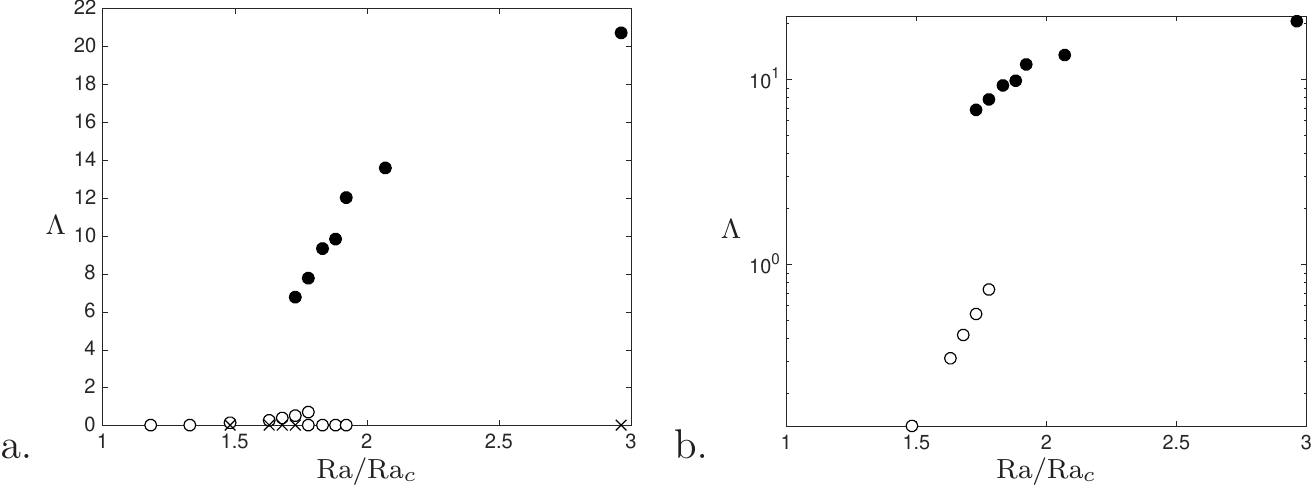}}
\caption{Weak-dipolar (circles) and strong-dipolar (bullets) branches 
for $\Ek=3 \times 10^{-4}$, $\Ek_\eta=1.7 \times 10^{-5}$, $\q=18 \, .$
Crosses indicate unstable solutions. 
{The Elsasser number is represented in a linear scale on panel (a) and a
  log-scale on panel (b), thus highlighting the lower branch.}}
\label{bif_strong}
\end{figure}

Recently, \citeasnoun{Dormy16} has shown that if $\Ek_\eta$ was small enough
in the numerical simulations, the viscous-dipolar mode could exhibit a
transition to a very different state, of stronger, yet still dipolar,
magnetic field. The corresponding bifurcation diagram is illustrated in
figure~\ref{bif_strong}. \citeasnoun{Dormy16} pointed that a ``cusp
catastrophe'' occurs in the bifurcation diagram as $\Ek_\eta$ is decreased
at fixed $\Ek \, .$ This catastrophe acounts for the transition between the
single branch reported on figure~\ref{fig_Morin}a, yet characterised by a sudden increase near
$\Ray/\Ray_c \simeq 3$, and the turning points highlighted by figure~\ref{bif_strong}.

The Elsasser number $\Lambda$ is usually assumed to be of order unity in
the strong-dipolar state. As can be seen in figure~\ref{bif_strong}, while
the values are ``of the order of unity'' (compared to the extreme values of
some parameters, such as those listed in (\ref{small_param})), they are
significantly greater than unity. For this reason a modified Elsasser
number
\begin{equation}
\Lambda^{\prime}= \Lambda \, {L}/(\Rm \, \ell^\star_B) \, ,
\end{equation} 
was used in \citeasnoun{Dormy16}, and was shown to offer a closer
measurement of the balance between the Coriolis and the Lorentz forces.
In the above definition, we used 
\begin{equation}
\Rm = \frac{U \, L}{\eta} \, , \qquad {\rm and} \qquad
{\ell_B} ^2 = \frac{\langle \mathbf{B}^2 \rangle}{\langle (\mathbf{\nabla} \times \mathbf{B})^2 \rangle} \, .
\end{equation} 
{Writting in dimensional form the ratio of the Lorentz to the Coriolis force yields
\begin{equation}
\frac{\{(\mu \rho)^{-1} \bfnabla \times \bfB \times \bfB\} }{\{2 \bfOmega \times \bfu\}}
= \frac{B^2}{2\Omega \mu \rho U \ell_B^\star} \, .
\end{equation}
The classical Elsasser number stems from $U\ell_B^\star \simeq \eta$, which is a sensible approximation in an 
asymptotic sense, as the magnetic Reynolds number is neither very large nor very small in this problem.
The modified Elsasser number offers a finer measure of this balance by writting $U\ell_B^\star / \eta \simeq 
\Rm \ell_B^\star / L \, .$
}

\citeasnoun{Dormy16} has shown that this dynamo state corresponds to
a primarily magnetostrophic balance, by comparing the radial
components of the curl of the Lorentz force and the Coriolis force (in
order to get rid both of the pressure gradient and of the buoyancy
force).

The transition from the viscous-dipolar to the strong-dipolar state,
characterised by a runaway field growth, as the turning point of the
weak-dipolar state is reached, was also reproduced in
\citeasnoun{Dormy16}.

\section{Geophysical relevance of the strong-dipolar state}

We discuss here additional simulations performed in the range $\Ek \in
[3\times10^{-4}, \, 10^{-5}]\, ,$ $\Ek_\eta \in [1.4\times10^{-6}, \,
  2.5\times10^{-5}]\, ,$ $\q \in [5,\, 18]\, $  {(see Table~1)}.
In these simulations
the parameters were chosen such that the strong-field state reported in the
previous section was maintained in the limiting process of decreasing both $\Ek$
and $\Ek_\eta$. This corresponds to a distinguished limit, relating the
small parameters, as introduced in \citeasnoun{Dormy16}.

\begin{figure}
\centerline{\includegraphics[width=1 \textwidth]{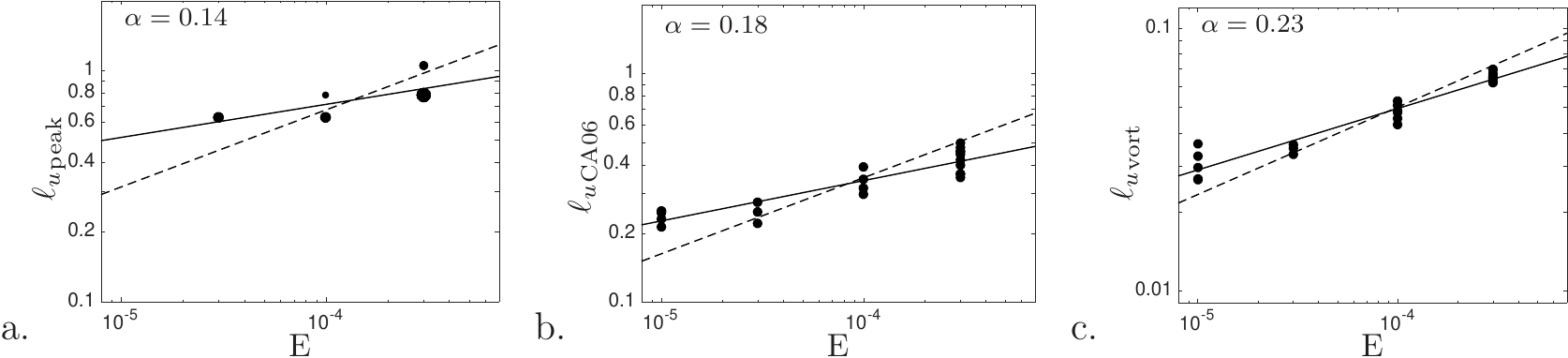}}
\caption{Evolution of the typical length scales of the flow as a function 
of the Ekman number, same representation as in figure~\ref{lu_weak}, but for numerical dynamos 
in the strong-dipolar state.}
\label{lu_strong}
\end{figure}

We can first {consider}, as we did for the first two dynamo states, the
length scales dependency with the Ekman number.  These are presented
in figure~\ref{lu_strong}. 
While viscosity clearly affects the small length scales of the flow, 
the length scale ${{\ell}_u}_{\rm peak}$ (corresponding to the energy spectrum peak)
appears reasonably independent on the Ekman number, consistent with a large-scale magnetostrophic 
balance.

\begin{figure}
\centerline{\includegraphics[width=1 \textwidth]{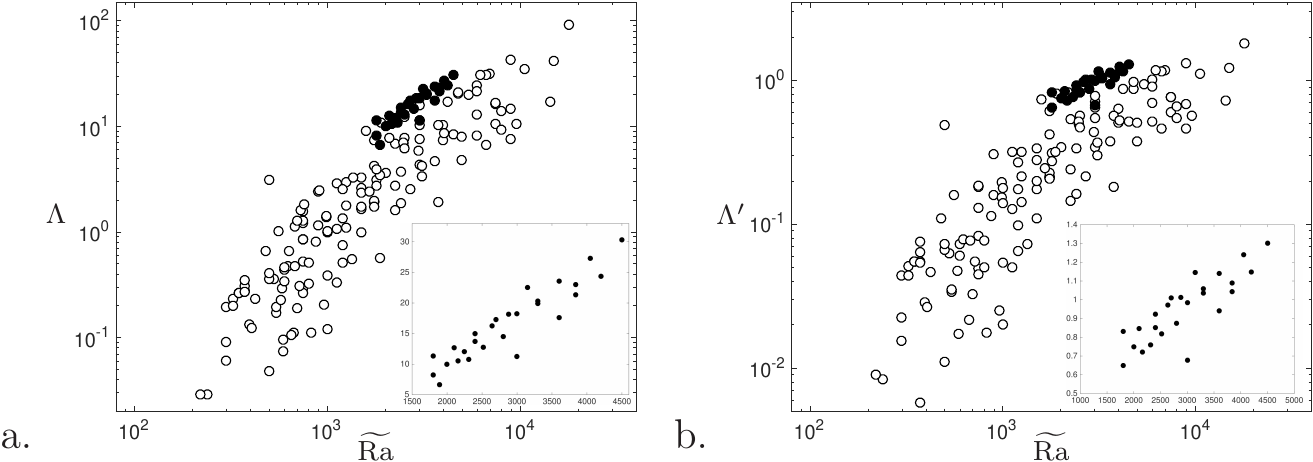}}
\caption{Strength of the magnetic field, as measured by (a) 
the Elsasser number $\Lambda \, ,$ (b)
the modified 
Elsasser number $\Lambda^\prime$, both as a function of the modified
Rayleigh number  
$\widetilde{\Ray}$, in the weak-dipolar regime (circles) and in the strong-dipolar regime (bullets). 
The insets present an enlarged representation of dynamos in the
strong-dipolar state.}
\label{Modif_Els}
\end{figure}

In order to assess that these simulations offer a sensible
approximation to the magnetostrophic limit, the Elsasser number
$\Lambda$ is reported in figure~\ref{Modif_Els}a as a function of the
combination $\widetilde{\Ray} = \Ray \,\q \, .$ We also report the modified
Elsasser number $\Lambda^\prime$ as a function of $\widetilde{\Ray}$ in
figure~\ref{Modif_Els}b. 
{As the field increases with $\Rm$ in these simulations, the modified Elsasser number 
presents a narrower range of variations than the Elsasser number.
The first important observation in the strong-dipolar state} is that indeed, whereas the Elsasser number
 is larger than unity and exhibits a clear
variation with $\widetilde{\Ray} \, ,$
the modified Elsasser number
is much closer to unity for all the simulations in the strong-dipolar state
(whereas a wide disparity can be observed in both plots 
for the weak-field state, {though with a 
narrower range in the case of the modified Elsasser number}).

The second essential information, is that despite the variations in $\Ek\, ,$ 
$\Ek_\eta \, ,$ and $\q \, $ all the strong-dipolar points appear to sit on a
single curve. They are only (or almost only) functions of
$\widetilde{\Ray}\, .$ This vindicates the scenario of \citeasnoun{Dormy16}
that these dynamos are approaching a dominant magnetostophic balance.

\begin{figure}
\centerline{\includegraphics[width=0.45 \textwidth]{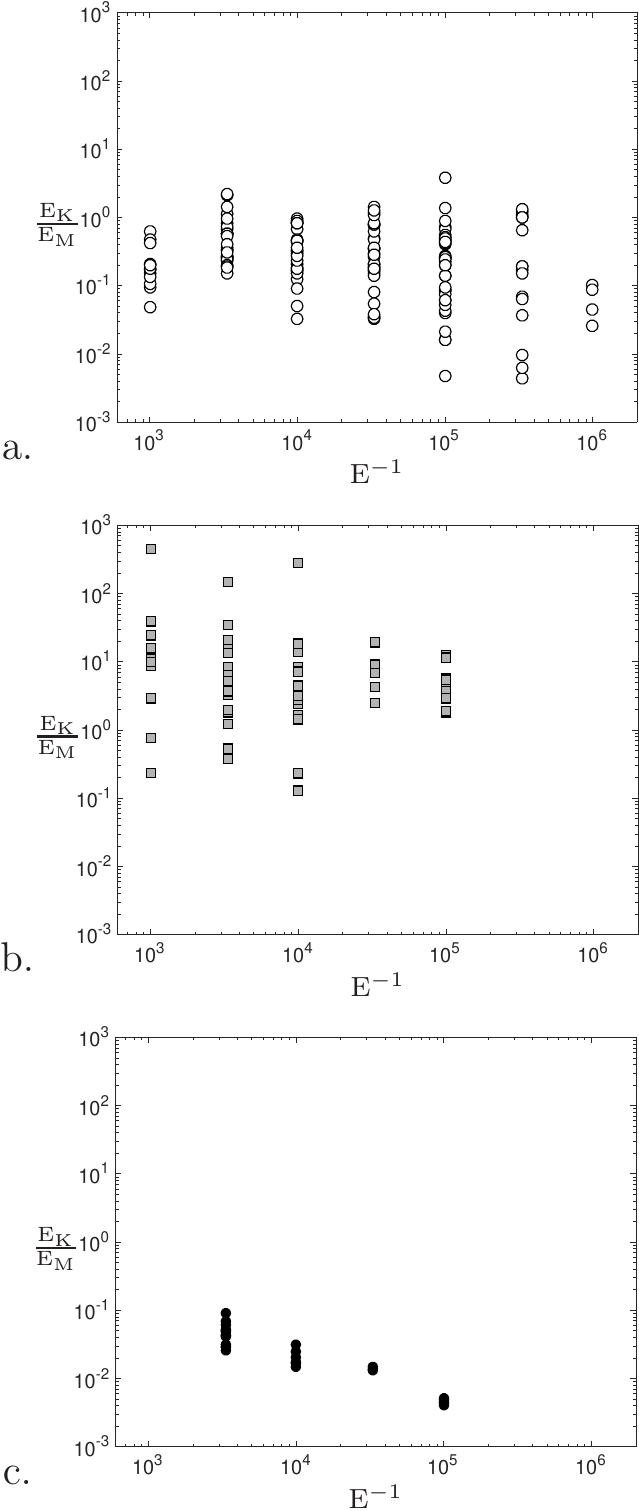}}
\caption{Ratio of the kinetic energy $E_{\rm K}$ over the magnetic energy
$E_{\rm M}$ as a function of the inverse Ekman number, for (a)
  weak-dipolar, 
(b) fluctuating-multipolar and (c) strong-dipolar dynamo states.}  
\label{FigEnergy}
\end{figure}

{
A key prediction on the strong-dipolar branch is that the kinetic
energy should be significantly lower than the magnetic energy}. This
contrasts with the strongly inertial regime in which equipartition
is eventually expected. Because of the smallness of the magnetic Ekman
number $\Ek_\eta$ in (\ref{eq_NS}), the magnetic energy $E_{\rm M}$
should here be much larger than the kinetic energy $E_{\rm K}$.  We
report in figure~\ref{FigEnergy} the evolution of the ratio of the
kinetic energy $E_{\rm K}$ over the magnetic energy $E_{\rm M}$ as a
function of the inverse Ekman number for the three dynamo states
discussed in this article.
This quantity varies significantly in the weak-dipolar state 
(see figure~\ref{FigEnergy}a). It is less than unity for most models, 
but no clear trend with the Ekman number can be emphasised.
In the multipolar-fluctuating state, the field is generally weaker, while 
the driving by buoyancy is stronger, as a result, most of these dynamos are
characterised by a ratio larger than unity (see figure~\ref{FigEnergy}b).
In figure~\ref{FigEnergy}c, the strong-dipolar models
exhibit a clear and 
systematic decrease of this ratio with the inverse Ekman number. 
They always correspond to lower values of this ratio than those achieved in
the weak-dipolar state.
For the smallest Ekman number considered here, this ratio reaches a value 
lower than on the two other branches.
This contrasts with \citeasnoun{Yadav16} who argue that largely
super-critical dynamos are needed to decrease this ratio (a result
which is however most likely correct on the weak-field branch).

\begin{figure}
\centerline{\includegraphics[width=1 \textwidth]{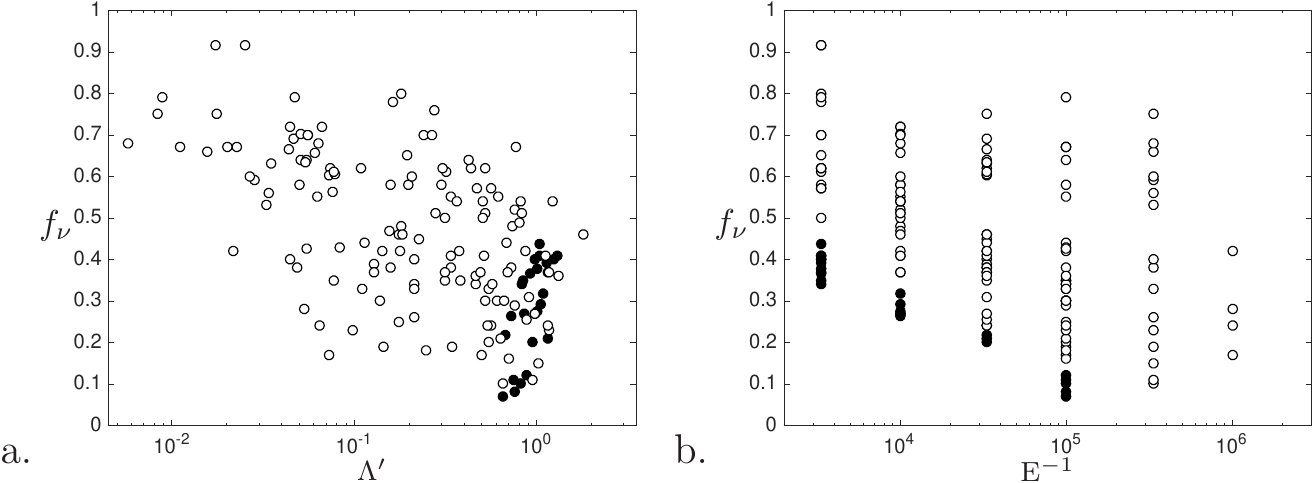}}
\caption{The $f_\nu$ parameter as a function of (a) the modified Elsasser
  number $\Lambda^{\prime}$ and (b) the inverse Ekman number $\Ek^{-1}$, 
  in the weak-dipolar state (circles) and in the strong-dipolar state (bullets).
}
\label{Visc_Diss}
\end{figure}

Of course, viscous forces are still present in the strong-dipolar state
simulations.  Their importance can be quantified by considering, as we
did for the weak and multipolar states, the fraction of viscous
dissipation. 
The ratio of the energy being dissipated by viscous forces to the
total energy dissipation (viscous and ohmic) should vanish as one
approaches the magnetostrophic limit.
Figure~\ref{Visc_Diss} presents the evolution of the ratio $f_\nu$ as a
function of various parameters.
Figure~\ref{Visc_Diss}a shows $f_\nu$ as a function of the modified
Elsasser number $\Lambda^\prime \, .$ It highlights that the strong-dipolar
state simulations span over a wide range of $f_\nu \, .$ While the viscous
dissipation never amounts to more than $50\%$ of the total dissipation,
this is a very significant variation.
It is then enlightening to represent $f_\nu$ as a function of the inverse
Ekman number. This graph, represented in figure~\ref{Visc_Diss}b, shows
that, for those dynamos that are in the strong-dipolar state, the fraction of
viscous dissipation indeed vanishes as the Ekman number decreases. So that
most of the energy is dissipated via ohmic dissipation for the smaller
Ekman numbers considered. 

{
The relation between figure~\ref{FigEnergy}c and figure~\ref{Visc_Diss} is subtle.
With our choice of non-dimensional form, the kinetic and magnetic energy are expressed,
in units of $\Omega \rho \eta L^3$, as
\begin{equation}
E_{\rm K} = \frac{\Ek_\eta}{2} \int \bfu ^2 \, \rmd V \, , \qquad
E_{\rm M} = \frac{1}{2} \int \bfB ^2 \, \rmd V \, .
\end{equation}
If our choice of units yields order one values for the integrals, then the ratio 
$E_{\rm K}/E_{\rm M}$ will vanish as $\Ek_\eta$. }

{
If we now form the energy equations, after integration by parts, we get
\begin{equation}
\frac{\rmd E_{\rm K}}{\rmd t} =  \int {\Ray}\,\q\, \T \, \bfu \cdot \bfr\, \rmd V
 - \int \bfnabla \times (\bfu \times \bfB) \cdot \bfB\, \rmd V
 - \Ek \, \int ( \bfnabla \times \bfu)^2 \, \rmd V \, ,
\label{dissip_vel}
\end{equation}
\begin{equation}
\frac{\rmd E_{\rm M}}{\rmd t} = \int \bfnabla \times (\bfu \times \bfB) \cdot \bfB \, \rmd V
- \int ( \bfnabla \times \bfB)^2 \, \rmd V\, ,
\label{dissip_mag}
\end{equation}
The above two equations can be rewritten in a condensed form
\begin{equation}
\frac{\rmd E_{\rm K}}{\rmd t} = P - L - \Ek D^u \, ,
\qquad
\frac{\rmd E_{\rm M}}{\rmd t} = L - D^B
\, .
\label{Energy_bal}
\end{equation}
The first term on the right of (\ref{dissip_vel}) is the energy
production term  $P$. In non-magnetic hydrodynamics, this energy has to be
entirely dissipated in a statistically steady state by the last lerm
on the right-hand-side of (\ref{dissip_vel}), i.e. the viscous
dissipation $\Ek D^u$. In magnetohydrodynamics, however, the second term on
the right-hand-side of (\ref{dissip_vel}), $L$, allows a transfer of energy
to the induction equation (it is equal and opposite to the first term
on the right of (\ref{dissip_mag})), where the energy can be ohmically dissipated 
via the last term in (\ref{dissip_mag}), $D^B$.
Equation~(\ref{Energy_bal}) stresses that $f_\nu = \Ek D^u / (\Ek D^u+D^B)$ vanishes 
in the limit of small Ekman numbers.
}

\section{Conclusion}
\begin{figure}
\centerline{\includegraphics[width=0.45 \textwidth]{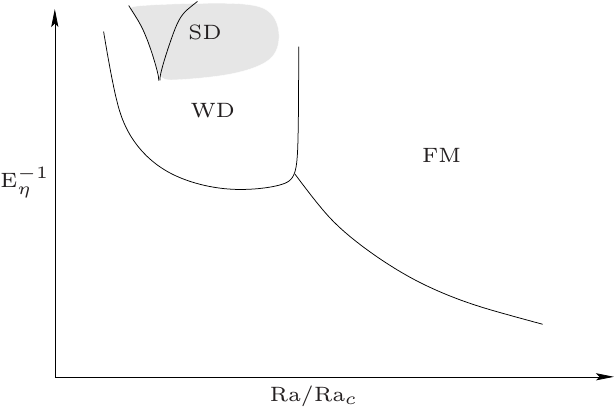}}
\caption{Phase-diagram illustrating, for a given Ekman number, the three
  different dynamo states observed in numerical simulations of the
  Geodynamo, namely the weak-dipolar (WD), the fluctuating-multipolar (FM)
  and the strong-dipolar (SD) states. The cusp indicates the region of
  bistability between the WD and the SD states. The grey shaded region
  marks the explored strong-dipolar state.}
\label{fig_state}
\end{figure}

We discussed the different dynamo states obtained in numerical models of
the Geodynamo. We show that at least three distinct states, characterised
by different forces balances can be highlighted in the available databases
of numerical dynamos. These are represented on a schematic phase-diagram at
fixed Ekman number on figure~\ref{fig_state}. The question of the most 
relevant choice of $\Ek_\eta$ at a given $\Ek$ is a difficult one. 
\citeasnoun{Dormy16} suggested that the ratio of these two small parameters 
should be determined through a distinguished limit, instead of 
systematically trying to maximise $\Ek_\eta$ at a given $\Ek$ (i.e. 
minimise $\Pm$).

{It should be noted that the distinguished limit advocated above
consists in relating the two small parameters $\Ek_\eta$ and $\Ek$ in the limiting process,
with the ratio $\Ek / \Ek_\eta \equiv \Pm$ vanishing in the limit. This constrasts with an alternative 
approach, which consists in dropping the inertial term, either entirely, or for the non-zonal terms only
\cite{Glatzmaier,JonesRoberts2000,Hughes16}, which amounts to an infinite $\Pm$ limit.}

The first dynamo state (WD), and probably the most documented one,
corresponds to a state in which viscosity is entering the main balance, and
influencing the convection length-scale. It does not mean to say that these
dynamos correspond to a pure ``VAC'' (Viscous-Archemedian-Coriolis)
balance. Other terms, such as inertial effects or Lorentz force, obviously
affect the solution. The second state (FM), originally highlighted by
\citeasnoun{Ku02}, corresponds to a state in which inertial forces became
strong enough so that the Rossby radius became comparable with the
convection length scale. In this state the dipolar component no longer
dominates the solution, and the field is referred to as multipolar.  The
third state (SD), introduced by \citeasnoun{Dormy16}, is characterised by a
runaway field growth from the weak state, due to a cusp catastrophe in the
bifurcation diagram. This state appears to correspond to the
magnetostrophic force balance anticipated in the Earth's core. Of course,
viscous effects, and inertial effects are still present, but they do not
enter the leading order balance. 

As the co-existence of the weak and the
strong-dipolar states for a given set of parameters (bistability) is
associated with the occurence of a fold in the bifurcation diagram
due to a cusp catastrophe, some models obtained at values of 
$\Pm = \Ek/\Ek_\eta$ below the appearance of this catastrophe will 
necessarily share some properties of the strong-dipolar state.

The strong-field state can only be achieved in numerical models so far by
adopting a large value of $\Pm = \Ek / \Ek_\eta \, .$ This is only due to
computational limitations, and this ratio can be decreased, following a
distinguished limit as advocated in \citeasnoun{Dormy16}.

Key issues remain, such as:
whether dynamo action can be observed for
$\Ray/\Ray_c$ lower than unity. This analytical prediction on the
strong-field branch has so far never been reproduced in numerical models
-- whether the point at which the cusp catastrophe occurs indeed decreases to
lower values of $\Pm=\Ek/\Ek_\eta$ as $\Ek$ is decreased -- whether a
transition from the strong-dipolar state to the fluctuating-multipolar
state can be observed for larger forcing.
Further studies will undoubtably be useful to address such open issues.

\appendix
\section{Numerical data}

\include{Table_strong}

\section*{Acknowledgements}
This work was granted access to the HPC resources of MesoPSL 
financed by the Region Ile-de-France and the project 
Equip@Meso (reference  ANR-10-EQPX-29-01) of the programme 
Investissements d'Avenir supervised by the Agence Nationale 
pour la Recherche. Numerical simulations were also carried  
out at CEMAG and TGCC computing centres (GENCI project x2013046698). 

\section*{References}
\bibliographystyle{jphysicsB}
\bibliography{biblio}

\end{document}

%% file: Table_strong.tex
\begin{table}
\caption{Direct Numerical Simulations performed in the strong-dipolar state.}
\centering
\begin{tabular}{rrrrrrrrrr}
\hline\\[-5mm]
$\widetilde{\Ray}$ & $\Nu$ &  ${\ell_u}_{\rm vort}$&${{\ell}_u}_{\rm CA06}$ & ${{\ell}_u}_{\rm peak}$ & $\ell_B$ &  $\Rm$ &  $\Lambda$ &  $E_{\rm K}/E_{\rm M}$ &   $f_\nu$\\
\hline
\multicolumn{10}{c}{$\Ek = 3\times10^{-4}$ \ \ \  $\Ek_\eta = 2.50\times10^{-5}$ \ \ \ $\q=12$}\\
1800 & 1.50 & 0.0696 & 0.4760 & 0.7854 & 0.0817 & 167 & 11.32 & 0.0312 & 0.34\\
2100 & 1.60 & 0.0700 & $-$ & $-$ & 0.0733 & 205 & 12.70 & 0.0410 & 0.35\\
2400 & 1.68 & 0.0674 & 0.4586 & 0.7854 & 0.0712 & 228 & 15.00 & 0.0433 & 0.37\\
2700 & 1.85 & 0.0647 & 0.4425 & 0.7854 & 0.0648 & 265 & 17.32 & 0.0503 & 0.38\\
3000 & 1.92 & 0.0651 & 0.3653 & 1.0472 & 0.0617 & 300 & 18.22 & 0.0617 & 0.40\\
3300 & 2.04 & 0.0638 & 0.3530 & 1.0472 & 0.0584 & 330 & 19.92 & 0.0691 & 0.41\\
3840 & 2.23 & 0.0622 & 0.3653 & 0.7854 & 0.0526 & 389 & 21.35 & 0.0900 & 0.44\\[2mm]
\multicolumn{10}{c}{$\Ek = 3\times10^{-4}$ \ \ \ $\Ek_\eta = 1.67\times10^{-5}$ \ \ \ $\q=18$}\\
1890 & 1.23 & $-$ & 0.4987 & $-$ & $-$ & 145 & 6.60 & 0.0262 & 1.00\\
2250 & 1.38 & $-$ & 0.4553 & $-$ & $-$ & 207 & 12.00 & 0.0289 & 1.00\\
3150 & 1.55 & 0.0655 & $-$ & $-$ & 0.0690 & 285 & 22.50 & 0.0295 & 0.37\\
3600 & 1.71 & 0.0648 & 0.4189 & $-$ & 0.0595 & 348 & 23.60 & 0.0428 & 0.39\\
4050 & 1.84 & 0.0632 & $-$ & $-$ & 0.0555 & 396 & 27.27 & 0.0479 & 0.40\\
4500 & 1.93 & 0.0622 & 0.3977 & 0.7854 & 0.0533 & 438 & 30.35 & 0.0527 & 0.41\\[2mm]
\multicolumn{10}{c}{$\Ek = 1\times10^{-4}$ \ \ \ $\Ek_\eta = 8.33\times10^{-6}$ \ \ \ $\q=12$}\\
2160 & 1.45 & 0.0528 & 0.3927 & 0.6283 & 0.0711 & 205 & 10.50 & 0.0165 & 0.26\\
2400 & 1.60 & 0.0508 & $-$ & $-$ & 0.0671 & 240 & 13.72 & 0.0175 & 0.27\\
2640 & 1.76 & 0.0429 & $-$ & $-$ & 0.0620 & 270 & 16.25 & 0.0148 & 0.27\\
2880 & 1.83 & 0.0487 & 0.3452 & 0.6283 & 0.0603 & 298 & 18.20 & 0.0206 & 0.28\\
3300 & 1.95 & 0.0478 & 0.3173 & 0.7854 & 0.0564 & 341 & 20.35 & 0.0248 & 0.29\\
3840 & 2.21 & 0.0455 & 0.2964 & 0.6283 & 0.0502 & 422 & 23.05 & 0.0319 & 0.32\\ [2mm]
\multicolumn{10}{c}{$\Ek = 3\times10^{-5}$ \ \ \ $\Ek_\eta = 3.00\times10^{-6}$ \ \ \ $\q=10$}\\
3000 & 1.81 & 0.0362 & 0.2732 & 0.6283 & 0.0496 & 335 & 11.25 & 0.0148 & 0.22\\
3600 & 2.13 & 0.0350 & 0.2493 & 0.6283 & 0.0474 & 394 & 17.60 & 0.0136 & 0.20\\
4200 & 2.40 & 0.0332 & 0.2212 & 0.6283 & 0.0467 & 455 & 24.40 & 0.0134 & 0.21\\[2mm]
\multicolumn{10}{c}{$\Ek = 1\times10^{-5}$ \ \ \ $\Ek_\eta = 2.00\times10^{-6}$ \ \ \ $\q=5$}\\
1800 & 2.00 & 0.0363 & 0.2513 & $-$ & 0.0638 & 198 & 8.20 & 0.0049 & 0.07\\
2000 & 2.17 & 0.0267 & 0.2513 & $-$ & 0.0632 & 211 & 10.00 & 0.0044 & 0.11\\[2mm]
\multicolumn{10}{c}{$\Ek = 1\times10^{-5}$ \ \ \ $\Ek_\eta = 1.43\times10^{-6}$ \ \ \ $\q=7$}\\
2310 & 1.90 & 0.0327 & 0.2474 & $-$ & 0.0579 & 246 & 10.80 & 0.0040 & 0.08\\
2520 & 2.12 & 0.0294 & 0.2310 & $-$ & 0.0546 & 285 & 12.75 & 0.0046 & 0.10\\
2800 & 2.30 & 0.0265 & 0.2137 & $-$ & 0.0518 & 320 & 14.50 & 0.0051 & 0.12\\[1mm]
\hline 
\end{tabular}
\label{tab:data}
\end{table}